%% file: main.tex
\documentclass[sigconf]{aamas}  

\usepackage{booktabs}

\usepackage{flushend}
\setcopyright{ifaamas}  
\acmDOI{doi}  
\acmISBN{}  
\acmConference[AAMAS'20]{Proc.\@ of the 19th International Conference on Autonomous Agents and Multiagent Systems (AAMAS 2020), B.~An, N.~Yorke-Smith, A.~El~Fallah~Seghrouchni, G.~Sukthankar (eds.)}{May 2020}{Auckland, New Zealand}  
\acmYear{2020}  
\copyrightyear{2020}  
\acmPrice{}  

\usepackage{pgfplots}
\usepackage[utf8]{inputenc}
\usepackage{etex}
\usepackage{booktabs}
\usepackage{amsmath,amssymb,amsfonts}
\usepackage{mathtools}
\usepackage{amsthm}
\usepackage{xcolor}
\usepackage{mathtools}
\usepackage{macros}
\usepackage{multirow}
\usepackage{multicol}
\usepackage{kantlipsum,cuted}
\usepackage{relsize}
\usepackage{mwe}
  
\everymath{ \displaystyle}
\RequirePackage{etex} 
\usepackage[ruled,vlined]{algorithm2e}
\usepackage{subcaption}

\usepackage{tikz}
\usetikzlibrary{positioning,calc,quotes}
\usepackage[T1]{fontenc}
\usepgfplotslibrary{fillbetween}
\usepackage{diagbox} 

\usetikzlibrary{arrows,automata,shapes,backgrounds}

\pgfplotsset{compat=1.9}
\RequirePackage{etex} 

\begin{document}

\title{Policy Synthesis for Factored MDPs with Graph Temporal Logic Specifications}  


\author{Murat Cubuktepe, Zhe Xu, Ufuk Topcu}

%
\affiliation{%
  \institution{The University of Texas at Austin}
  \city{Austin} 
  \state{Texas} 
}
\email{mcubuktepe@utexas.edu, zhexu@utexas.edu, utopcu@utexas.edu}
%
%
%
%
%
%
%

\begin{abstract}  
We study the synthesis of policies for multi-agent systems to implement \emph{spatial-temporal} tasks. 
We formalize the problem as a \textit{factored} Markov decision process subject to so-called \emph{graph temporal logic} specifications. 
The transition function and the spatial-temporal task of each agent depend on the agent itself and its neighboring agents.
The structure in the model and the specifications enable to develop a distributed algorithm that, given a factored Markov decision process and a graph temporal logic formula, decomposes the synthesis problem into a set of smaller synthesis problems, one for each agent.
We prove that the algorithm runs in time linear in the total number of agents.
The size of the synthesis problem for each agent is exponential only in the number of neighboring agents, which is typically much smaller than the number of agents.
We demonstrate the algorithm in case studies on disease control and urban security. 
The numerical examples show that the algorithm can scale to hundreds of agents.
\end{abstract}


\maketitle


\section{Introduction}

\begin{figure}[t]
	\centering
	\begin{tikzpicture}[innerblue/.style={minimum size=0.6cm,circle,draw=blue!40,fill=blue!40,thick,scale=0.4},inner/.style={minimum size=0.6cm,circle,draw=blue!50,fill=blue!50,thick,scale=0.4},innergreen/.style={minimum size=0.6cm,circle,draw=purple!50!black,fill=purple!50!black,thick,scale=0.4},innerred/.style={minimum size=0.6cm,circle,draw=orange!60!black,fill=orange!60!black,thick,scale=0.4},innerred1/.style={minimum size=0.6cm,circle,draw=red!20,fill=red!20,thick,scale=0.4}]
  \tikzstyle{every state}=[minimum size=1.3cm, fill=none,node distance=1.8cm,semithick,font=\large]
  \tikzstyle{every node}=[fill=none,semithick,font=\large]
\node[innergreen] (s11) at (-0.3,0.3) {};
\node[innerred] (s12) at (-0.3,-0.3) {};
\node[innerblue] (s13) at (0.3,-0.3) {};
\draw[->,semithick,red] (s11) -- (s12) {};
\draw[->,semithick,red,bend left] (s12) to (s11) {};
\draw[->,semithick,red] (s11) -- (s13) {};
\draw[->,semithick,red] (s12) -- (s13) {};

\node[innergreen] (s21) at (1.5,0.3) {};
\node[innerred] (s22) at (1.5,-0.3) {};
\node[innerblue] (s23) at (2.1,-0.3) {};
\node[innerred] (s24) at (2.1,0.3) {};

\draw[->,semithick,red] (s21) -- (s22) {};
\draw[->,semithick,red] (s23) -- (s21) {};
\draw[->,semithick,red] (s22) -- (s23) {};
\draw[->,semithick,red] (s21) -- (s24) {};
\draw[->,semithick,red] (s24) -- (s23) {};

\node[innerred] (s31) at (-0.3,-1.5) {};
\node[innerblue] (s32) at (-0.3,-2.1) {};
\node[innergreen] (s33) at (0.3,-2.1) {};
\node[innerred] (s34) at (0.3,-1.5) {};

\draw[->,semithick,red] (s32) -- (s31) {};
\draw[->,semithick,red] (s31) -- (s33) {};
\draw[->,semithick,red] (s33) -- (s32) {};
\draw[->,semithick,red] (s34) -- (s31) {};
\draw[->,semithick,red] (s33) -- (s34) {};

\node[innerred] (s41) at (1.5,-1.5) {};
\node[innerblue] (s42) at (1.5,-2.1) {};
\node[innergreen] (s43) at (2.1,-2.1) {};
\draw[->,semithick,red] (s42) -- (s41) {};
\draw[->,semithick,red] (s43) -- (s41) {};
\draw[->,semithick,red] (s42) -- (s43) {};
\draw[->,semithick,red,bend left] (s41) to (s43) {};

  \node[state] (1)                    {};
  \node[state]         (2) [right of=1] {}; 
  \node[state]         (3) [below of=1] {}; 
  \node[state]         (4) [right of=3] {};;
  \draw[line width=2pt,green] (1) -- (2); 
  \draw[line width=2pt,green] (2) -- (4);
  \draw[line width=2pt,green] (2) -- (3);
  \draw[line width=2pt,green] (3) -- (1);
  \draw[line width=2pt,green] (4) -- (3);
  \node [draw=none] at (0.8,-0.5) {${}$};
  \node [draw=none] at (2.6,0.2) {${}$};
  \node [draw=none] at (3.4,-0.6) {${}$};
  \node [draw=none] at (-0.9,-0.3) {$1$};
  \node [draw=none] at (2.8,-0.3) {$2$};
  \node [draw=none] at (-0.9,-2.1) {$3$};
  \node [draw=none] at (2.8,-2.1) {$4$};
  \node [draw=none] at (-1.35,0.3) {};
  \node [draw=none] at (2.8,0.3) {};
  \node [draw=none] at (-0.9,-1.5) {};
  \node [draw=none] at (2.8,-1.5) {};
\end{tikzpicture}
\caption{An example of an undirected graph. The nodes ($1, 2, 3,$ and $4$) of the graph indicate the agents. Each agent has its own set of states and actions. We draw an edge between two agents if they share the current state information. The color of the states gives the label for each agent.} 
	\label{graph_exp}
\end{figure}
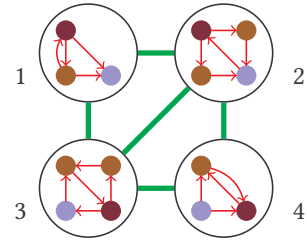

We consider a system where multiple agents coordinate to achieve a set of \emph{spatial-temporal tasks} defined over an underlying graph modeling the interaction between the agents.
For example, in the graph in Figure~\ref{graph_exp}, each node represents an agent, and each agent has its own set of states and actions. 
We draw edges between \emph{neighboring} agents that exchange their current state information. 
For each agent, the transition function between its states also depends on the current state of the neighboring agents.
The labels at each node of the graph provide information about the agents related to the task.
The spatial-temporal task specified for each agent depends both on the agent itself and its neighboring agents.
For example, the different nodes in a graph as shown in Figure~\ref{graph_exp} can model different police officers.
The states of each node can represent the intersections that the corresponding police officer is monitoring in a city.
An edge between two nodes exists if the two corresponding police officers can share their state information.
The task might be \textquotedblleft agent $3$ or a neighboring agent of $3$ should be in the intersection labeled as blue in every two time steps\textquotedblright.


We model the behavior of each agent as a Markov decision process (MDP)~\cite{puterman2014markov}, which has been widely used to model and solve sequential decision-making problems.
We can represent the states of the overall multi-agent system by explicitly enumerating all possible states of the agents, meaning each agent is interacting with all other agents.
The resulting state space of the composed MDP will scale exponentially in the number of agents, and the representation will be impractical for policy synthesis.
The interaction between the agents is typically \emph{sparse}, and each agent in the system exchanges their current state information with only a few other agents. 
Examples of such systems appear in biochemical networks~\cite{prescott2014layered}, smart grids~\cite{dorfler2014sparsity}, swarm robots~\cite{pickem2017robotarium}, and disease control~\cite{choisy2007probabilistic}. 
 
Because of the sparsity of the interaction between agents, the composed MDP is typically a factored MDP~\cite{guestrin2003efficient,cheng2013variational,guestrin2002multiagent,forsell2006approximate}, which provides an efficient representation for MDPs with multiple agents. 
In factored MDPs, the transition probabilities of an agent often depend only on a small number of other agents that are neighbors in the underlying graph. 
Such a representation alleviates the need for explicitly enumerating all possible states of the agents, and facilitates the synthesis of policies for systems with a large number of agents.

Related work in factored MDPs typically focuses on defining a reward function and computing a policy to maximize the expected reward.
A wide range of tasks, such as avoiding certain parts of an environment, cannot be expressed in any reward function~\cite{littman2017environment,hahn2019omega}.
Linear temporal logic~\cite{Pnueli,BK08} is a language that can concisely express such tasks. 
However, linear temporal logic cannot express spatial-temporal tasks involving multiple agents and is limited to single-agent tasks.

Our first contribution is to represent spatial-temporal tasks in a specification language called graph temporal logic (GTL)~\cite{zhe_GTL}.
GTL formulas represent tasks such as \textquotedblleft the police officer at node $3$ or their neighboring officers should visit intersection labeled as blue in every two hours\textquotedblright.
GTL is an extension of linear temporal logic and focuses on the spatial-temporal properties of the labels on a graph. 
We use GTL formulas to express tasks that concern a set of agents on the graph.

Expressing the spatial-temporal tasks as GTL specifications gives us the following advantages. 1) GTL provides a concise way to constrain the behavior of the agents to satisfy the spatial-temporal tasks, which cannot be done by designing a reward function in general. 2) GTL resembles natural languages, thus requirements specified by practitioners can be translated into GTL specifications. 3) GTL specifications can be converted into a deterministic finite automaton~\cite{zhe_GTL}, which facilitates the policy synthesis utilizing the techniques in formal methods.


Our second contribution is to develop a centralized algorithm for the factored MDP policy synthesis problem subject to GTL specifications.
We first derive the centralized algorithm based on a linear programming problem (LP).
The proposed LP is a generalization of the LP formulation for MDPs subject to linear temporal logic specifications~\cite{BK08,forejt2011quantitative}.
By solving the LP, we synthesize a policy for each agent, which is a function of the states of the agent and its neighboring agents on the graph. 

Our third contribution is to develop a distributed algorithm, since the centralized algorithm may not scale to a large number of agents. 
We develop the distributed algorithm by extending the centralized synthesis algorithm.
Specifically, the algorithm decomposes the centralized problem into smaller synthesis problems, one for each agent. 
The algorithm then solves the smaller synthesis problems in parallel at each iteration.
The running time of the algorithm is linear with the number of agents in the graph and exponential with the number of neighbors for each agent, which is typically much smaller due to the sparsity of the graph.

We demonstrate the effectiveness of the algorithm on two examples with a large number of agents. 
In the first example, we consider disease management in crop fields with GTL specifications~\cite{sabbadin2012framework}.
A contaminated crop field can infect its neighbors, and the yield of that crop field decreases.
The GTL specifications ensure that some of the crop fields are treated immediately to prevent the spreading of disease.
We consider maximizing the expected yield of the crop fields while ensuring that some of the crop fields are treated immediately.
We also show that the decentralized algorithm outperforms the centralized algorithm with a large number of agents. 
Specifically, the running time of the decentralized algorithm scales linearly with the number of agents.
In the second example, we consider an urban security problem. 
The objective of the problem is to assign patrol tasks to police officers such that certain critical locations in the city are sufficiently monitored.
We express the task of monitoring the critical locations in GTL specifications.
The police officers need to coordinate with each other to satisfy the GTL specifications.
The results show that the proposed distributed algorithm scales to hundreds of agents while ensuring that the agents achieve tasks that are specified in GTL formulas.

\noindent\textbf{Related work.} Related work on factored MDPs considers maximizing an expected reward (or, minimizing an expected cost). The existing results consider optimizing for a value function using approximate linear
programming~\cite{guestrin2003efficient,guestrin2002multiagent,forsell2006approximate}, approximate policy iteration~\cite{sabbadin2012framework,peyrard2006mean}, and approximate value iteration~\cite{guestrin2003efficient,cheng2013variational}.
However, optimizing for a value function may not be sufficient to ensure safety or performance guarantees in time-related tasks that include multiple agents. 
Reference~\cite{hahn2019omega} shows that no reward structure, in general, is not sufficient to capture tasks that are given by temporal logic specifications.

The work in~\cite{sahin2017provably} considers the problem of coordinating multiple homogeneous agents subject to constraints on the number of agents achieving a task given in temporal logic specifications. 
However, they do not allow agents to be heterogeneous and do not differentiate between different agents on different tasks. 
Recent work in~\cite{haghighi2016robotic} proposes a spatial-temporal logic for swarm robots. 
The proposed logic is less expressive than GTL, and their solution approach involves solving a mixed-integer linear program, which is significantly more challenging--in theory and in practice--than the optimization problems that arise in the proposed approach. Reference~\cite{verginis2018motion} proposes a framework for potential-based collision avoidance of multi-agent systems with temporal logic specifications. 
However, the proposed controller is centralized, and the controller does not scale to a large number of agents.

Related work on multi-agent partially observable MDPs (POMDPs) exploited similar graph-based interactions between agents. The work in~\cite{nair2005networked} exploits the graph structure between the agents and synthesizes a policy that maximizes an objective. \cite{kumar2016dual} describes a dual mixed-integer linear program to synthesize \emph{finite-state controllers} for decentralized POMDPs by generalizing the LP formulation for MDPs that we also utilize. A dynamic-programming based approach is used in~\cite{seuken2007memory} to compute finite-length policies for decentralized POMDPs. \cite{seuken2007memory} uses heuristics to improve the scalability of the exact dynamic-programming based methods~\cite{hansen2004dynamic}, which can only synthesize policies for up to a handful time steps.

 
\section{Preliminaries}

A \emph{probability distribution} over a finite set $\distDom$ is a function $\distFunc\colon\distDom\rightarrow\Ireal$ with $\sum_{\distDomElem\in\distDom}\distFunc(\distDomElem)=\distFunc(\distDom)=1$. 
The set $\distDom$ of all distributions is $\Distr(\distDom)$. For two sets $A, B$ we define $A \subseteq B$ if $B$ contains all elements of $A$.

\begin{definition}[Undirected graph]
Let $G=(V, E)$ be an undirected graph, where $V=\lbrace 1,\ldots,M \rbrace$ is a finite set of agents and $E$ is a finite set of edges. We use $e=\{1, 2\}$ to denote that the edge $e\in E$ connects $1$ and $2\in V$. For agents $i, j$, we call $j$ a neighboring agent of $i$, if there is an edge $e$ that connects $i$ and $j$. Let $N(i)\subseteq V$ be the set of agent $i$ and the neighboring agents of the agent $i$. We denote $N(i,j)=N(i)\cap N(j)$ and $N(i\setminus j)=N(i)\setminus N(j)$. The number of agents in $V$ is given by $M=|V|.$  
\end{definition}


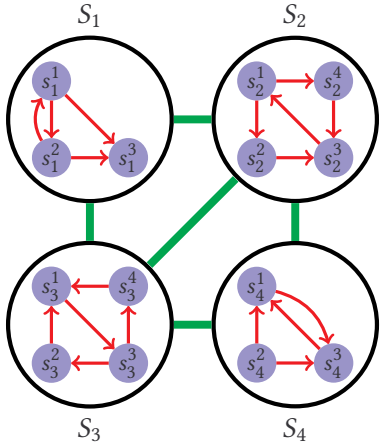
\begin{figure}[t]
    \centering

\begin{tikzpicture}[remember picture,
  inner/.style={minimum size=0.15cm,circle,draw=blue!40,fill=blue!40,ultra thick,inner sep=0.0pt},
  outer/.style={minimum size=0.2cm,draw=black,fill=white!20,ultra thick,inner sep=0.0pt},
  blank/.style={draw=none,fill=none,ultra thick,inner sep=0.001pt}
  ]
  \node[circle,outer,draw=black,label={\Large $S_1$}] (A) { 
    \begin{tikzpicture}
      \node [inner] (ai)  {$s^1_1$};
      \node [inner,below=0.5 cm of ai] (aii) {$s^2_1$};
      \node [inner,right=0.5 cm of aii] (aiii) {$s^3_1$};
      \draw[->,red,very thick] (ai) -- (aii); 
      \draw[->,red,very thick] (ai) -- (aiii);
      \draw[->,red,very thick] (aii) -- (aiii);
      \draw[->,red,very thick,bend left] (aii) to (ai);

    \end{tikzpicture}
  };
  \node[circle,outer,draw=black,right=0.5 cm of A,label={\Large $S_2$}] (B) {
    \begin{tikzpicture}
      \node [inner] (bi)  {$s^1_2$};
      \node [inner,below=0.5 cm of bi] (bii) {$s^2_2$};
      \node [inner,right=0.5 cm of bii] (biii) {$s^3_2$};
      \node [inner,right=0.5 cm of bi] (biv) {$s^4_2$};
      \draw[->,red,very thick] (bi) -- (bii);
      \draw[->,red,very thick] (biii) -- (bi);
      \draw[->,red,very thick] (bii)-- (biii);
      \draw[->,red,very thick]  (bi) -- (biv);
      \draw[->,red,very thick]  (biv) -- (biii);    

    \end{tikzpicture}
  };
    \node[circle,outer,draw=black,below=0.5 cm of A,label={}] (C) {
    \begin{tikzpicture}
      \node [inner] (ci)  {$s^1_3$};

      \node [inner,below=0.5 cm of ci] (cii) {$s^2_3$};
      \node [inner,right=0.5 cm of cii] (ciii) {$s^3_3$};
      \node [inner,right=0.5 cm of ci] (civ) {$s^4_3$};
      \draw[->,red,very thick] (cii) -- (ci);
      \draw[->,red,very thick] (ci) -- (ciii);
      \draw[->,red,very thick] (ciii) -- (cii);
      \draw[->,red,very thick]  (civ)-- (ci);
      \draw[->,red,very thick] (ciii) -- (civ);

    \end{tikzpicture}
  };
    \node[circle,outer,draw=black,right=0.5 cm of C,label={}]  (D) {
    \begin{tikzpicture}
      \node [inner] (di)  {$s^1_4$};

      \node [inner,below=0.5 cm of di] (dii) {$s^2_4$};
      \node [inner,right=0.5 cm of dii] (diii) {$s^3_4$};
      \draw[->,red,very thick] (dii) -- (di);
      \draw[->,red,very thick] (diii) -- (di);
      \draw[->,red,very thick] (dii) -- (diii);
      \draw[->,red,very thick,bend left] (di) to (diii);
    \end{tikzpicture}
  }; 
          \node [blank,below=0.1cm of C] (CC)  {\Large $S_3$};
        \node [blank,below=0.1 of D] (DD)  {\Large $S_4$};

   \draw[green,line width=3pt] (A) --  (B);
   \draw[green,line width=3pt] (A) --  (C);
   \draw[green,line width=3pt] (B) --  (C);
   \draw[green,line width=3pt] (B) --  (D);
   \draw[green,line width=3pt] (C) --  (D);
\end{tikzpicture}
    \caption{An example of a factored MDP. For each agent $i$ in $V$, $S_i$ depicts the state space of the agent $i$, and $s_i^j$ depicts the state $j$ of the agent $i$. The arrows between the states of the agent $i$ shows the transitions between states of the agent $v$. 
    For example, at a certain time step $t$, let the current states be $s^1_1$, $s^2_2$, and $s^4_3$ for the agents $1, 2,$ and $3$ respectively. Then, the transition probability of agent $1$ at state $s^1_1$ to $s^2_1$ and $s^3_1$ is a function of $s^1_1$, $s^2_2$, and $s^4_3$.}
        \label{fig:graph_mdp}
\end{figure}

\subsection*{Factored MDP}

\begin{definition}[Factored MDP]
A factored Markov decision process (MDP) $\MdpInit$ on the undirected graph $G=(V, E)$ is defined by a finite set $S$ of states, which is given as a Cartesian product of the states for each agent $i$ in the graph, i.e., $S= S_1 \times S_2 \times  \cdots \times S_M$, an initial state $\sinit \in S$, a finite set $\Act=\Act_1 \times \Act_2 \times \cdots \times \Act_M$ of actions, a transition function $\probmdp =\probmdp_1 \times \probmdp_2 \times \cdots \times \probmdp_M$, where for each agent $i$,  $\probmdp_i\colon S_{N(i)}\times\Act_{N(i)}\rightarrow\Distr(S_{N(i)})$ gives the transition function for agent $i$, where $S_{N(i)}$ and $\Act_{N(i)}$ denote the Cartesian product of the sets of states and actions of the agents in $N(i)$ respectively, a finite set $\pi$ of atomic propositions, a labeling function $\Label = \Label_1 \times \Label_2 \times \cdots \times \Label_M$, where $\Label_i \colon S_{N(i)} \rightarrow 2^{\pi}$ that labels each state $s \in S_{N(i)}$ with a subset of atomic propositions $\Label_i (s) \subseteq \pi$ and a reward function $R= R_1 \times R_2 \times \cdots \times R_M$, where $R_i \colon S_{N(i)}\times\Act_{N(i)}\rightarrow\R_{\geq 0}$ assigns a reward to state-action pairs for agent $i$. We use $s(t)$ to denote the states of all agents in $V$ at time index $t$.
\end{definition}

\noindent We give two examples to illustrate the concepts related to factored MDPs. The first example shows the relationship between the agents in the graph. The second example illustrates how the transition probabilities between states of an agent depend on the neighboring agents.

\begin{figure}[t]
    \centering

    \begin{tikzpicture}[  inner/.style={minimum size=0.17cm,circle,draw=blue!40,fill=blue!40,ultra thick,inner sep=3.5pt},  blank/.style={draw=none,fill=none,ultra thick,inner sep=0.001pt},scale=4.5]
      \node [inner] (ai)  {\Large $s^1_1$};
      \node [inner,below=0.6cm of ai] (aii) {\Large $s^2_1$};
      \node [inner,right=0.6cm of aii] (aiii) {\Large $s^3_1$};
      \draw[->,red,very thick] (ai) -- (aii); 
      \draw[->,red,very thick,bend left] (aii) to (ai); 
      \draw[->,red,very thick] (ai) -- (aiii);
      \draw[->, red,very thick] (aii) -- (aiii);
      \node[blank,below left=0.3cm and 0.2cm of ai] {\Large $\probmdp_1(s^1_1|s^2_1,s^2_2,s^4_3,\act^1_1)$};
      \node[blank,below right=-0.2cm and 0.3cm of ai] {\Large $\probmdp_1(s^3_1|s^1_1,s^2_2,s^4_3,\act^1_1)$};
     \node[blank,below right=0.2cm and -0.7cm of aii] {\Large $\probmdp_1(s^3_1|s^2_1,s^2_2,s^4_3,\act^1_1)$};

    \end{tikzpicture}

    \caption{An example of transition probabilities between states for the agent $1$ on the underlying graph of a factored MDP if the state in the neighboring agents are $s=2$ for the agent $2$ and $s=4$ for the agent $3$ if an action $\act=1$ is taken for the agent $1$.}
        \label{fig:graph_mdp2}
\end{figure}
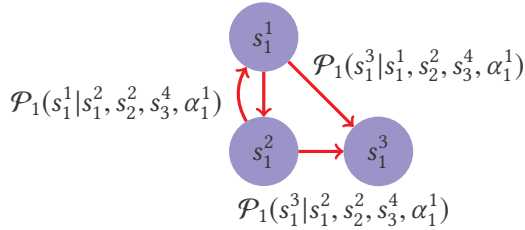

\begin{example}
    We show an example of a factored MDP in Figure~\ref{fig:graph_mdp}. Consider a graph $G=(V,E)$, where $V=\lbrace 1, 2, 3, 4\rbrace$ and $E=\lbrace \lbrace 1, 2\rbrace,\lbrace 1, 3\rbrace,\lbrace 2, 3 \rbrace, \lbrace 2, 4\rbrace, \lbrace 3, 4 \rbrace \rbrace$. The black nodes in Figure~\ref{fig:graph_mdp} represent the agents in the graph $G$. We denote the edges between the agents in the graph with green lines in Figure~\ref{fig:graph_mdp}. For $i \in V$, we denote the state space and action space of the agent $i$ as $S_i$ and $\Act_i$, and we denote the state $j$ in $S_i$ as $s^j_i$. We treat $s^j_i$ and $s^l_k$ to be different states if $i\neq k$ or $j\neq l$. For this example, $N(1)=\lbrace 1,2,3\rbrace, N(4)=\lbrace 2,3,4\rbrace$, and $N(1,4)=\lbrace 2,3\rbrace.$
    \end{example}
    
\begin{example}
    We show an example of the transition probabilities between states of the agent $1$ in a factored MDP in Figure~\ref{fig:graph_mdp2}. 
    For this example, we assume that $S_{N(1)}=\lbrace  s_1,s_2,s_3\rbrace$, for $s_1 \in S_1, s_2 \in S_2, s_3 \in S_3$ and $\Act_{N(1)}=\Act_1$, meaning the transition probability of the agent $1$ is a function of the states of agent $1$ and the neighboring agents, and the action of agent $1$. 
    The transition probabilities between the states of $S_1$ are given with red lines in Figure~\ref{fig:graph_mdp2}. 
    For example, in Figure~\ref{fig:graph_mdp2}, the transition probabilities between states in $S_1$ for a given action $\act^1_1$ is a function of $s_1 \in S_1$, $s_2 \in S_2$ and $s_3 \in S_3$.  
\end{example}

\begin{definition}[Policy]\label{def:scheduler} 
	A \emph{memoryless and randomized policy} for a factored MDP $\mdp$ is a function $\sched = \sched_1 \times \sched_2 \times \cdots \times \sched_M$, where for each agent $i$, $\sched_i \colon  S_{N(i)} \rightarrow\Distr(\Act_{N(i)})$.
The set of all policies over $\mdp$ is $\Sched_\mdp$.
\end{definition}
Applying a policy $\sched\in\Sched^\mdp$ to a factored MDP $\mdp$ yields an \emph{induced factored Markov chain} $\mdp_\sched$. 

\begin{definition}[Factored induced MC]\label{def:induced_dtmc}
	For a factored MDP $\MdpInit$ and a policy $\sched\in\Sched_{\mdp}$, the factored \emph{MC induced by $\mdp$ and $\sched$} is $\mdp_\sched=(S, \sinit, \Act, \probmdp_\sched, \pi, \Label, R)$, where
	\begin{align*}
	\displaystyle	\probmdp_\sched(s' | s)=\sum_{\act\in\Act(s)} \sched(s,\act)\cdot\probmdp(s' | s,\act) \;\forall s,s'\in S.
	\end{align*} 
\end{definition}

\begin{definition}[Trajectory]\label{def:trajectory}
A finite or an infinite sequence $\varrho_{\sched} = s(0)s(1)s(2)\ldots$ of states generated in $\mathcal{M}$ under a policy $\sched$ $\in$ $\Sched_{\mathcal{M}}$ is called a \textit{trajectory}. The state $s(0)$ denotes the initial state $\sinit$.
\end{definition}



Given an induced factored MC $\mathcal{M}_{\sched}$, starting from the initial state $\sinit$, the state visited at step $t$ is given by a random variable $X(t)$. The probability of reaching state $s'$ from state $s$ in one step, denoted $\mathbb{P}(X(t+1)=s' | X(t) = s)$ is equal to $\mathcal{P}_{\sched}(s' | s)$.  We can extend one-step reachability over a set of trajectories $\varrho_{\sched}$, i.e., $\mathbb{P}(X(n)=s(n), \allowbreak\ldots, X(0)=s(0) )=\mathbb{P}(X(n) = s(n) | X(n-1) = s(n-1) )\cdot \mathbb{P}(X(0)=s(0), \ldots, X(n-1)= s(n-1))$. We denote the set of all trajectories in $\mathcal{M}$ under the policy $\sched$ by $Tr_{\sched}(\mathcal{M})$.

\subsection*{Graph Temporal Logic}

We review the theoretical framework of graph temporal logic (GTL) introduced in \cite{zhe_GTL}.    


We denote $Y$ as the set of states for the edges, which is given as a Cartesian product of the states for each edge $e_i$, i.e., $Y= Y_1 \times Y_2 \times  \cdots \times Y_{|E|}$, where $Y_i$ is the state space of the edge $e_i$. Let $\mathbb{T}=\{1, 2, \dots\}$ be a discrete set of time indices.  We use $y(t)$ to denote the states of all edges in $E$ at time index $t$.

\begin{definition}[Graph-temporal trajectory]~\cite{zhe_GTL}
	\label{graph}
	A \textit{graph-temporal trajectory} on a graph $G$ is a tuple $g=(s, y)$, where $s:\mathbb{T}\rightarrow S$ assigns a node label for each node $i\in V$ at each time index $t\in\mathbb{T}$, and $y:\mathbb{T}\rightarrow Y$ assigns a edge label for each edge $e_i\in E$ at each time index $t\in\mathbb{T}$. 
\end{definition}

\begin{definition}[Node and edge propositions]~\cite{zhe_GTL}
	\label{node}
	An \textit{atomic node proposition} is a Boolean valued map on $S$, which is a predicate. An \textit{edge proposition} is a Boolean valued map on $Y$.
\end{definition}

\begin{figure}[t]
    \centering

    \begin{tikzpicture}[  inner/.style={minimum size=0.17cm,circle,draw=black!40,fill=white!40,ultra thick,inner sep=3.5pt},  blank/.style={draw=none,fill=none,ultra thick,inner sep=0.001pt},scale=4.5]
      \node [inner] (v1)  {\Large $1$};
      \node [inner,below=1.2cm of v1] (v2) {\Large $2$};
      \node [inner,right=2.2cm of v1] (v3) {\Large $3$};
      \node [inner,right=2.2cm of v2] (v4) {\Large $4$};
      \draw[very thick] (v1) -- (v2); 
      \draw[very thick] (v1) -- (v3); 
      \draw[very thick] (v2) -- (v4); 
      \draw[very thick] (v2) -- (v3); 
      \draw[very thick] (v4) -- (v3);       
      \node[blank,above=0.15cm of v1] {\Large ${\color{red}\lbrace \textrm{blue, red} \rbrace}$};
      \node[blank,below=0.15cm of v2] {\Large ${\color{red}\lbrace \textrm{orange, blue} \rbrace}$};
      \node[blank,above=0.15cm of v3] {\Large ${\color{red}\lbrace \textrm{red, orange} \rbrace}$};
      \node[blank,below=0.15cm of v4] {\Large ${\color{red}\lbrace \textrm{blue, red} \rbrace}$};
      \node[blank,below left=0.5cm and 0.0cm of v1] {\Large $e_1$};
      \node[blank,below left=0.8cm and -0.25cm of v1] {\Large $\color{blue}\lbrace 1,2 \rbrace$};
      \node[blank,above right=-0.15cm and 0.85cm of v1] {\Large $\color{blue}\lbrace 4,1 \rbrace$};
      \node[blank,above right=0.25cm and 1.10cm of v1] {\Large $e_2$};
    \node[blank,below right=-0.15cm and 1.1cm of v2] {\Large $e_3$};
      \node[blank,below right=0.1cm and 0.85cm of v2] {\Large $\color{blue}\lbrace 1,3 \rbrace$};
      \node[blank,below right=0.50cm and 2.95cm of v1] {\Large $e_4$};
      \node[blank,below right=0.80cm and 2.65cm of v1] {\Large $\color{blue}\lbrace 2,3 \rbrace$};
    \node[blank,below left=0.45cm and 1.35cm of v3] {\Large $\color{blue}\lbrace 3,2 \rbrace$};
      \node[blank,below left=0.15cm and 1.65cm of v3] {\Large $e_5$};
    \end{tikzpicture}

    \caption{An example of a graph-temporal trajectory of time length $2$ on an undirected graph, with the red items indicating node labels, and the blue numbers indicating edge labels. The first item in the labels is for time index $1$, and the second item in the labels is for time index $2$.}
\label{fig:gtl_example}
\end{figure}
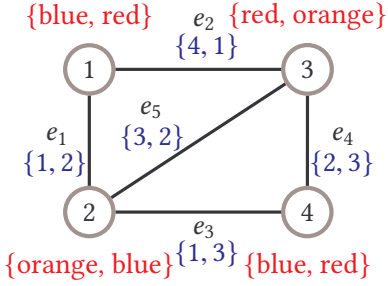

Let $\pi$ be an atomic node proposition, and $\mathcal{O}(\pi)$ be the subset of $S$ for which the atomic proposition $\pi$ is true. Similarly, let $\rho$ be an edge proposition, and $\mathcal{O}(\rho)$ be subset of $Y$ for which $\rho$ is true.

A graph-temporal trajectory $g=(s, y)$ satisfies an atomic node proposition $\pi$ at a node $v$ and at a time index $t$, which we denote as $(g,v,t)\models\pi$, if and only if $s_v(t) \in \mathcal{O}(\pi)$.
A graph-temporal trajectory $g=(s, y)$ satisfies an edge proposition $\rho$ at an edge $e$ and at a time index $t$, which we denote as $(g,e,t)\models\rho$, if and only if $y_e(t) \in \mathcal{O}(\rho)$.

\begin{example}
Referring back to the example in Figure~1, we show an example with $4$ police officers. The node and edge labels are from a graph-temporal trajectory $g$ for a time index $1$ and $2$. The node labels represent the intersection that the corresponding police officer is monitoring, and the edge labels represent the distance between two neighboring officers. The atomic node proposition $ \pi=(x = \textrm{blue} \vee \textrm{orange})$ is satisfied by $g$ at nodes $1$, $2$, and $4$ at time index $1$, and at nodes $2$, and $3$ at time index $2$. The edge proposition $\rho = (y \leq 2)$ is satisfied by $g$ at $e_1$, $e_3$, and $e_4$ at time index $1$, and at $e_1$, $e_2$, and $e_5$ at time index $2$.
\end{example}



\begin{definition}[Neighboring operation]~\cite{zhe_GTL}
	\label{N}
	Given a graph-temporal trajectory $g=(s,y)$ on a graph $G$, an edge proposition $\rho$, and a subset $V'\subseteq V$ of nodes, we define the neighboring operation $\bigcirc_{\rho}$ : $2^{V}\times\mathbb{T}\rightarrow 2^{V}\times\mathbb{T}$ as~\cite{zhe_GTL}
	\[
	\begin{split}
	&\bigcirc_{\rho}(V',t) =\big(\{v | \exists v'\in V', \exists e\in E, e=\{v',v\}, (g,e,t)\models\rho\},t\big).
	\end{split}
	\]
	$\bigcirc_{\rho}(V',t)$ are the set of nodes that can be reached from $V'$ through an edge $e$ if the edge proposition $\rho$ is true by graph-temporal trajectory $g$ at time index $t$. We can apply neighboring operations successively. 	
\end{definition}

\begin{example}
For the graph-temporal trajectory $g$ on the graph $G$ at time index $1$ in Figure~\ref{fig:gtl_example},
\begin{align*}
    \bigcirc_{y \leq 2}(\lbrace{1\rbrace},1)&=(\lbrace{2\rbrace},1),\\
      \bigcirc_{y \leq 2}  \bigcirc_{y \leq 2}(\lbrace{1\rbrace},1)&=\bigcirc_{y \leq 2}(\lbrace{2\rbrace},1)=(\lbrace{1, 4\rbrace},1).\\
\end{align*}
\end{example}


We define the syntax of a GTL formula $\varphi$ recursively as~\cite{pLTL2014,zhe_GTL}
\begin{align}\nonumber
\begin{split}
	&\varphi:=
	\pi~|~\exists^{N}(\bigcirc_{\rho_{n}}\cdots \bigcirc_{\rho_{1}})\varphi~|~\neg\varphi~|~X\varphi~|~\varphi\wedge\varphi~|~\varphi\mathcal{U}\varphi~|~\Diamond_{\sim i}\varphi, 
\end{split}
\end{align}
where $n$ and $N$ are positive integers, $\pi$ is an atomic         
node proposition, $\rho_j$ for $j=1,\dots,n$ are edge propositions. The formula $\exists^{N}(\bigcirc_{\rho_{n}}\cdots \bigcirc_{\rho_{1}})\varphi$ reads as \textquotedblleft there exists at least $N$ nodes under the neighboring operation $\bigcirc_{\rho_{n}}\cdots \bigcirc_{\rho_{1}}$ that satisfy the formula $\varphi$\textquotedblright. $\lnot$ denotes negation of a formula, and $\wedge$ denotes conjunction of two formulas. $X$ and $\mathcal{U}$ are temporal operators, where $X$ represents \textquotedblleft next\textquotedblright~and $\mathcal{U}$ represents \textquotedblleft until\textquotedblright. $\Diamond_{\sim i}$ is a parametrized temporal operator representing \textquotedblleft parametrized eventually\textquotedblright, where $\sim\in\{\ge,\le\}$, which denotes that the formula should be satisfied within or after $i$ time steps. Many common operators that can be found in linear temporal logic such as $\vee$ (disjunction), $\Diamond$ (eventually), $\Box$ (always), $\Box_{\sim i}$ (parametrized always), $\mathcal{U}_{\sim i}$ (parametrized until) and $\Rightarrow$ (implication) can be derived from the above-mentioned operators.


We define the satisfaction relation $(g, v, t)\models\varphi$ for a graph-temporal trajectory $g$ at node $v$ and at time index $t$ with respect to a GTL formula $\varphi$ as~\cite{pLTL2014,zhe_GTL}                          
\[
\begin{split}
&(g, v, t)\models\pi\quad\mbox{iff}\quad s_v(t)\in\mathcal{O}(\pi)\\
&(g, v, t)\models\exists^{N}(\bigcirc_{\rho_{n}}\cdots \bigcirc_{\rho_{1}})\varphi~\mbox{ iff } \exists \lbrace{ i_1,\ldots,i_N\rbrace}\in V~(i_j \neq i_k \text{ for } j\neq k),\\
&\text{ such that } \forall j\in\{1,\dots, N\}, i_j\in\bigcirc_{\rho_{n}}\cdots \bigcirc_{\rho_{1}}(v,t)~ \mbox{and}~ (g, i_j, t)\models\varphi                
\end{split}                                             
\]      
\[
\begin{split}
(g, v, t)\models\lnot\varphi\quad\mbox{iff}\quad & (g, v, t)\not\models\varphi\\
(g, v, t)\models X\varphi\quad\mbox{iff}\quad& (g, v, t+1)\models\varphi\\
(g, v, t)\models\varphi_{1}\wedge\varphi_{2}\quad\mbox{iff}\quad & (g, v, t)\models\varphi_{1}~\mbox{and}~(g, v, t)\models\varphi_{2}\\
(g, v, t)\models\varphi_{1}\mathcal{U}\varphi_{2}\quad\mbox{iff}\quad &  \exists
t'\ge t, \mbox{s.t.}~(g, v, t')\models\varphi_{2},\\
&  (g, v, t^{\prime\prime})\models\varphi_{1}, \forall t^{\prime\prime}\in[t, t']\\
(g, v, t)\models\Diamond_{\sim i}\varphi\quad\mbox{iff}\quad & \exists
t'\sim t+i, \mbox{ such that }~(g, v, t')\models\varphi
\end{split}
\]

A graph-temporal trajectory $g$ at a node $v$ at a time index $t$ satisfies the formula $\exists^{N}(\bigcirc_{\rho_{n}}\cdots \bigcirc_{\rho_{1}})\varphi$ if there exist at least $N$ nodes in $(\bigcirc_{\rho_{n}}\cdots \bigcirc_{\rho_{1}})(v,t)$ where the formula $\varphi$ is satisfied by a graph-temporal trajectory $g$ at time index $t$. 
We also define that a graph-temporal trajectory $g$ satisfies a GTL formula $\varphi$ at a node $v$, denoted as $(g, v)\models\varphi$, if $g$ satisfies $\varphi$ at node $v$ at time index 0.
  
For the GTL formulas that we consider, specifically \textit{syntactically co-safe (resp. safe)} formulas, we can build a DFA $\mathcal{A}^{\varphi,v}$ (resp. $\mathcal{A}^{\lnot\varphi,v}$) over $\mathcal{AP}$ that accepts precisely the graph-temporal trajectories that satisfy (resp. violate) the GTL formula $\varphi$ at node $v$~\cite{zhe_GTL}.   
We focus on syntactically co-safe and syntactically safe formulas for the remainder of the paper, which can be satisfied or violated with a graph-temporal trajectory of finite time length.

For simplicity, we focus on the node labels only in this paper. In this case, the graph-temporal trajectory $g$ reduces to a trajectory $\varrho$ (as defined in Definition \ref{def:trajectory}). We use $(\varrho, v , t)\models \varphi$ to denote the fact that a trajectory $\varrho$ satisfies the GTL formula $\varphi$ at node $v$ at time index $t$.

\section{Problem Formulation}\label{GTL_sec}

To solve the factored MDP policy synthesis problem, we synthesize a policy for each agent that satisfies the GTL specification. We construct a factored MDP that captures all trajectories of a factored MDP $\mdp$ satisfying a GTL formula $\varphi$ by taking the product of $\mdp$ and the DFA $\mathcal{A}^{\varphi}$ corresponding to the GTL formula $\varphi$. We represent the GTL specifications as reachability specifications on the product factored MDP. 




\begin{definition}[Product factored MDP]
Let $\MdpInit$ be a factored MDP and $\DraInit$ be a DFA. 
The product factored MDP is a tuple $\ProductInit$ with a finite set $\Sp=S \times Q$ of states, an initial state $\sinitp=(\sinit,q)\in \Sp$ that satisfies $q=\delta(\qinit,\mathcal{L}(\sinitp))$, a finite set $\Act$ of actions, a labeling function $\Labelp((s,q))=\{q\}$, a transition function $\probmdpp_i((s,q), \act, (s',q'))= \probmdp_i(s,\act,s')\;  \text{if} \quad q'=\delta(q,\mathcal{L}(s'_i))$, and  $\probmdpp_i((s,q), \act, (s',q'))=0 \; \allowbreak \text{otherwise}$, a reward function $\Rp$ that satisfies $\Rp(s,q,\act)=R(s,\act)$ for $s \in S$ and $\act \in \Act$, the acceptance condition $\Accp=\{A_1^\mathrm{p},\ldots,A_k^\mathrm{p} \}$, where $A_i^\mathrm{p}=\Sp\times A_i$ for all $A_i\in \Acc$. 
We assume all accepting states are absorbing.
\end{definition}

A memoryless and randomized policy for a product factored MDP $\mdpp$ is a function $\schedp = \schedp_1 \times \schedp_2 \times \cdots \times \schedp_M$, where $\schedp_i \colon  \Sp_{N(i)} \rightarrow \Distr(\Act_{N(i)})$. A memoryless policy $\schedp$ is a finite-memory policy $\sched'$ in the underlying factored MDP $\mdpp$. Given a state $(s,q) \in \Sp$, we consider $q$ to be a memory state and define $\sched'(s)=\schedp(s,q)$.

%
 
Let $\schedp\in\Sched_{\mdpp}$ be a policy for $\mdpp$ and let $\sched'\in\Sched_{\mdp}$ be the policy on $\mdp$ constructed from $\schedp$ through the procedure explained above. 
The trajectories of the factored MDP $\mdp$ under the policy $\sched'$ satisfy the GTL specification $\varphi$ with a probability of at least $\lambda$ at node $v$ and at a time index $t$, i.e., $\mathbb{P}_{\mdp_{\sched'}}((\varrho_{\sched'},v,t)\models \varphi)\geq \lambda$, if and only if the trajectories of the product factored MDP $\mdpp$ under the policy $\schedp$ reach some accepting states in $\mdpp$ with a probability of at least $\lambda$~\cite{BK08}.

\begin{definition}[Occupancy measure]
The occupancy measure $o^\sched_i$ of a policy $\sched_i$ for a set of neighboring agents $N(i)$ of a factored MDP $\mdp$ is defined as 
\begin{align}
&   \displaystyle o^{\sched}_i(\hat{s}_{N(i)},\hat{\act}_{N(i)})=
\mathbb{E}\Big[ \sum_{t=0}^{\infty}\mathbb{P}(\hat{s}^{N(i)}(t)=\hat{s}_{N(i)},\nonumber\\
&\hat{\act}_{N(i)}(t)=\hat{\act}_{N(i)}|s_{N(i)}(t)=\sinit_{N(i)})\Big],
\end{align}
where $\hat{s}_{N(i)}(t)=\lbrace s_{j_1}(t), \ldots ,s_{j_{|N(i)|}}(t) \rbrace \in S_{N(i)}$ and $\hat{\act}_{N(i)}(t) \allowbreak= \allowbreak \lbrace  \act_{j_1}(t), \allowbreak\ldots, \act_{j_{|N(i)|}}(t) \rbrace \allowbreak\in \Act_{N(i)}$ denote the state and action of the agent $i$ and the neighboring agents in $\mdp$ at time index $t$. 
The equality $\hat{s}_{N(i)}(t)=\hat{s}_{N(i)}$ means all elements of $\hat{s}_{N(i)}(t)$ and $\hat{s}_{N(i)}$ are the same. The occupancy measure $o^{\sched}_i(\hat{s}_{N(i)},\hat{\act}_{N(i)})$ is the expected number of times to take the action ${\act}_{j_k}$ at the state ${s}_{j_k}$ for all $k \in N(i)$ under the policy $\sched_i$.
\end{definition}



\subsection*{Problem Statement} 

In this section, we formally state the policy synthesis problem of a factored MDP subject to GTL specifications. 
Our objective is to synthesize a policy that induces a stochastic process with a maximum expected reward whose trajectories satisfy the given GTL specification with at least a desired probability. 
To this end, we synthesize a policy $\schedp \in \Sched_{\mdpp}$ that reaches and stays in the accepting states in $\mdpp$ with probability of at least $\lambda$. 

\begin{problem}
Given a factored product MDP $\mdpp$, $\lambda$, $k$, and a set of agents $\hat{V},$ compute a policy $\schedp \in \Sched_{\mdpp}$ that solves the problem
\begin{align}
    & \displaystyle \underset{\sched^\mathrm{p} \in \Sched_{\mdpp}}{\textnormal{maximize}}\quad \mathbb{E}\left[ \sum_{t=0}^{\infty}R(s(t), \act(t))\right]\\
    & \displaystyle \textnormal{subject to} \quad \mathbb{P}_{\mdpp_{\schedp}}( (\varrho_{\schedp},v,k) \models \varphi)\geq \lambda,
\end{align}
where $\mathbb{P}_{\mdpp_{\schedp}}((\varrho_{\schedp},v,k)\models \varphi)$ denotes the probability of satisfying the GTL specification $\varphi$ with the trajectory $\varrho_{\schedp}$ for some agents $v \in \hat{V}$. $\hat{V}$ is the set of agents that have a GTL specification that is required to be satisfied at the time index $k$ in the factored product MDP $\mdpp$ under the policy $\schedp$. Without loss of generality, we assume $k=1.$
\end{problem}

\section{Policy Synthesis}

We now describe the proposed approach to synthesize a policy for each agent to solve Problem~1. We first give a centralized formulation based on a linear programming problem. We then develop a distributed approach based on the centralized formulation. 

\subsection{Centralized Approach}

In this section, we propose a linear programming problem (LP) for solving Problem~1. 
Our solution is based on the multi-objective dual LP formulation to compute a policy that maximizes the expected reward while satisfying a graph temporal logic specification $\varphi$ in a factored MDP with $|V|=1$, which is an MDP~\cite{puterman2014markov,forejt2011quantitative}.
Let $\bar{S}$ be the set of all states in the MDP $\mdp$ that are not in $Acc$. Then, we define the variables of the dual LP formulation as follows.
		\begin{itemize}
			\item $o(s,\act)\in [0,\infty)$ for each state $s\in \bar{S}$ and action $\act\in\Act$ defines the occupancy measure of a state-action pair for the policy $\sched$, i.e., the expected number of times of taking action $\act$ in state $s$. 
			\item $o(s) \in [0, 1]$ for each state $s \in Acc$ defines the occupancy measure of an accepting state $s \in Acc$, which is equal to the probability of reaching the accepting state $s$.	\end{itemize}
Note that the variables $o(s)$ are defined in the interval $[0, 1]$ instead of $[0, \infty)$, as they represent the probability of reaching an accepting state $s \in Acc$. The variable $o(s,\act)$ represents the expected number of taking action $\act$ in a non-accepting state $s$, thus it can exceed $1$. We refer to~\cite{forejt2011quantitative,savas2019entropy} for further explanation of the domain of the variables.

The dual LP is given by
\begin{align}
		   \displaystyle	&\text{maximize} \quad   \displaystyle \sum_{s\in \bar{S}}o(s)R(s)\label{eq:policylp:obj}\\
			&\text{subject to} \nonumber
\\
		& 	\forall s\in \bar{S},
		\quad     \displaystyle	\sum\limits_{\act\in\Act}o(s,\act) = \sum\limits_{s'\in \bar{S}}\sum\limits_{\act\in\Act}\probmdp(s|s',\act)o(s',\act)+\mu(s),	\label{eq:policylp:welldefined_sched}\\
				&	\forall s\in Acc, \quad 
				\displaystyle 	\displaystyle o(s)= \sum_{s'\in \bar{S}}\sum_{\act\in\Act}\probmdp(s|s',\act)o(s',\act)+\mu(s),\label{eq:policylp:mec_sched}\\
		   \displaystyle		& \displaystyle 	 \sum_{s \in Acc} o(s) \geq \lambda,\label{eq:policylp:probthresh}
\end{align}
where $\mu(s)=1$ if $s=\sinit$ and $\mu(s)=0$ if $s\neq \sinit$. 
The constraints~\eqref{eq:policylp:welldefined_sched} and~\eqref{eq:policylp:mec_sched} ensure that the expected number of times transitioning to a state $s \in \bar{S}$ is equal to the expected number of times to take action $\act$ that transitions to a different state $s' \in S$. These constraints are also referred to as \emph{flow} constraints~\cite{puterman2014markov}.
The constraint~\eqref{eq:policylp:probthresh} ensures that the specification $\varphi$ is satisfied with a probability of at least $\lambda$.

For any optimal solution $o$ to the LP in~\eqref{eq:policylp:obj}--\eqref{eq:policylp:probthresh}, 
\begin{align}
   \displaystyle \sched(s,\act)= \dfrac{o(s,\act)}{   \sum_{\act'\in\Act}o(s,\act')}\label{eq:occupmeasure}
\end{align} 
is an optimal policy, and $o$ is the occupancy measure of $\sched$, see~\cite{puterman2014markov} and~\cite{forejt2011quantitative} for details. 

\subsection{LP-based Policy Synthesis of Factored MDPs}

We now describe our centralized approach for the policy synthesis problem for factored MDPs subject to GTL specifications. 
Let $\bar{S}^{\mathrm{p}}_{N(i)}$ be the set of all states in the product factored MDP $\mdpp$ that are not in $A^\mathrm{p}_i$ for each agent $i$. 
Then, we define the variables of the LP for policy synthesis as follows.
		\begin{itemize}
			\item $o_{i}(\hat{s}_{N(i)},\hat{\act}_{N(i)})\in [0,\infty)$ for each set of neighboring states $\hat{s}_{N(i)} \in \bar{S}^{\mathrm{p}}_{N(i)}$ and actions $\hat{\act}_{N(i)}\in\Act_{N(i)}$ defines the occupancy measure of a state-action pair for $\sched_i$.
			\item $o_{i}(\hat{s}_{N(i)}) \in [0, 1]$ for each state $\hat{s}_{N(i)} \in A^p_i$ defines the probability of reaching an accepting state $s\in A^p_i$.
		\end{itemize}

The objective of the LP is given by 

\begin{align}
\text{maximize} \quad    \displaystyle& \displaystyle \sum\limits_{i=1}^{M}  \sum\limits_{\hat{s}_{N(i)}\in \bar{S}^{\mathrm{p}}_{N(i)}}\sum\limits_{\hat{\act}_{N(i)}\in\Act_{N(i)}}\nonumber\\ & \displaystyle o_i(\hat{s}_{N(i)},\hat{\act}_{N(i)})R_i(\hat{s}_{N(i)},\hat{\act}_{N(i)}).\label{eq:graph_mdp_obj}
\end{align}

For each agent $i \in V$, and state $\hat{s}_{N(i)} \in \bar{S}^{\mathrm{p}}_{N(i)}$, the constraints 
\begin{align}
    & \displaystyle \sum\limits_{\hat{\act}_{N(i)}\in\Act_{N(i)}}o_i(\hat{s}_{N(i)},\hat{\act}_{N(i)}) -\mu(\hat{s}_{N(i)})=\label{eq:graph_mdp_flow}\\
    &\displaystyle \sum\limits_{\hat{s}'_{N(i)}\in \bar{S}^{\mathrm{p}}_{N(i)}}\sum\limits_{\hat{\act}_{N(i)}\in\Act_{N(i)}}\probmdp_i^{\mathrm{p}}(\hat{s}_{N(i)}|\hat{s}'_{N(i)},\hat{\act}_{N(i)})o_i(s'_i,\hat{\act}_{N(i)})\nonumber
\end{align}
denote the flow constraints, similar to the constraints~\eqref{eq:policylp:welldefined_sched}.

For each agent $i \in V$, and state $\hat{s}_{N(i)} \in A^p_i$, the constraints

\begin{align}
    & \displaystyle o_i(\hat{s}_{N(i)}) -\mu(\hat{s}_{N(i)})=\label{eq:graph_mdp_flow2}\\
    &\displaystyle\sum\limits_{\hat{s}'_{N(i)}\in \bar{S}^{\mathrm{p}}_{N(i)}}\sum\limits_{\hat{\act}_{N(i)}\in\Act_{N(i)}}\probmdp_i^{\mathrm{p}}(\hat{s}_{N(i)}|\hat{s}'_{N(i)},\hat{\act}_{N(i)})o_i(s'_i,\hat{\act}_{N(i)})\nonumber
\end{align}
\noindent denote the flow constraints for the accepting states, analogous to the constraints~\eqref{eq:policylp:mec_sched}.

For agents $i, j \in V$ such that $N(i)\cap N(j) \neq \emptyset$, we ensure that the occupancy measure is consistent in the states and actions of agents $k \in N(i,j)$. 
Thus, the agents take account of its neighboring agents' occupation measures during the policy computation. 
For each set of states $\hat{s}_{N(i,j)} \in S_{N(i,j)}$ and actions $\hat{\act}_{N(i,j)} \in \Act_{N(i,j)}$, the constraints
\begin{align}
     & \displaystyle \sum\limits_{\hat{s}_{N(i)}\supseteq \hat{s}_{N(i,j)}}\sum\limits_{\hat{\act}_{N(i)}\supseteq \hat{\act}_{N(i,j)}}o_i(\hat{s}_{N(i)},\hat{\act}_{N(i)}) =\nonumber\\
     &  \displaystyle \sum\limits_{\hat{s}_{N(j)}\supseteq \hat{s}_{N(i,j)}}\sum\limits_{\hat{\act}_{N(j)}\supseteq \hat{\act}_{N(i,j)}}o_j(\hat{s}_{N(j)},\hat{\act}_{N(j)}) \label{eq:graph_mdp_neighbor}
\end{align}
ensure that the time spent in the set of states $\hat{s}_{N(i,j)}$ and taking the set of actions $\hat{\act}_{N(i,j)}$ is equal for the policies of agents $i$ and $j$.

Finally, the constraints
\begin{align}
    \quad & \displaystyle  \sum_{\hat{s}_{N(i)} \in A^p_i} o_i(\hat{s}_{N(i)}) \geq \lambda_i  \label{eq:graph_mdp_spec}  
    \end{align} 
 encode the specification constraints for each agent $i \in V$, similar to the constraints~\eqref{eq:policylp:probthresh}.

We illustrate the constraints~\eqref{eq:graph_mdp_neighbor} by an example.
    \begin{example}
    Consider the factored MDP in Figure~\ref{fig:graph_mdp} and the agents $1$ and $2$. $N(1,2)=\lbrace1,2,3\rbrace, N(1\setminus 2)=\lbrace 3\rbrace$ and $N(2 \setminus 1)=\lbrace 4 \rbrace$. Therefore, to ensure that the occupancy measure is consistent for agents $1$ and $2$, we add the constraints
    \begin{align}
        & \displaystyle \sum\limits_{s_3 \in S_{N(1\setminus 2)}}\sum\limits_{\act_3 \in \Act_{N(1\setminus 2)}}o_1(\hat{s}_{N(1,2)},s_3,\hat{\act}_{N(1,2)},\act_3)=\nonumber\\
       & \displaystyle  \sum\limits_{s_4 \in S_{N(2\setminus 1)}}\sum\limits_{\act_4 \in \Act_{N(2\setminus 1)}}o_2(\hat{s}_{N(1,2)},s_4,\hat{\act}_{N(1,2)},\act_4).
    \end{align}
for $\hat{s}_{N(1,2)} \in S_{N(1,2)}$ and $\hat{\act}_{N(1,2)} \in \Act_{N(1,2)}$.

\end{example}
    
The LP, which is given by the objective~\eqref{eq:graph_mdp_obj} and the constraints~\eqref{eq:graph_mdp_flow}--\eqref{eq:graph_mdp_spec}, computes a policy for each agent $i$ that satisfies the GTL specification and maximizes the expected reward. However, the LP in~\eqref{eq:graph_mdp_obj}--\eqref{eq:graph_mdp_spec} can be time consuming to solve if the number of agents $M$ is large. In the next section, we propose a distributed approach that runs in time linear in $M$.
    
    \subsection{Distributed Approach}

In this section, we discuss how we can solve the LP in~\eqref{eq:graph_mdp_obj}--\eqref{eq:graph_mdp_spec} in a distributed manner. We utilize alternating direction method of multipliers (ADMM)~\cite{gabay1975dual,boyd2011distributed} to solve a large-scale factored MDP synthesis problem by decomposing them into a set of smaller problems. The iterations for ADMM does not necessarily converge to an optimal solution with $M>2$ agents~\cite{chen2016direct}. Therefore, we pose the multi-block problem into an equivalent two-block problem, and apply the \emph{primal-splitting ADMM} algorithm~\cite{wang2013solving} to the equivalent problem to ensure convergence.



\subsubsection{Primal-Splitting ADMM}

The LP in~\eqref{eq:graph_mdp_obj}--\eqref{eq:graph_mdp_spec} with $M$ agents can be written as following optimization problem
\begin{align}
\text{minimize} & \displaystyle\quad ~\sum_{i=1}^M f_i(o_i)\label{eq:multi_block_obj} \\
\text{subject to} &\displaystyle\quad ~\sum_{i=1}^M A_io_i =0,\label{eq:multi_block_cons}
\end{align}
where $f_i(o_i)$ is the negative of the objective in~\eqref{eq:graph_mdp_obj} and encodes the constraints in~\eqref{eq:graph_mdp_flow}--\eqref{eq:graph_mdp_flow2} and \eqref{eq:graph_mdp_spec} for each agent $i$. The constraints in~\eqref{eq:multi_block_cons} depict the constraints in~\eqref{eq:graph_mdp_neighbor} in a compact form. The matrices $A_i$ encodes the coefficients in~\eqref{eq:graph_mdp_neighbor}. Specifically, for agent $i$, the objective $f_i(o_i)$ in~\eqref{eq:multi_block_obj} can be expressed as following

\begin{align}
    &\text{minimize}  \quad f_i(o_i) = (g_1(o_i)+\mathcal{I}_0(g_2(o_i))+\mathcal{I}_0(g_3(o_i))+\mathcal{I}_+(g_4(o_i)))
    \end{align}
where $g_1(o_1)$ is the negative of the objective in~\eqref{eq:graph_mdp_obj}, and $g_2(o_i)=$ 
    \begin{align*}
         &  \displaystyle \sum\limits_{\hat{\act}_{N(i)}\in\Act_{N(i)}}o_i(\hat{s}_{N(i)},\hat{\act}_{N(i)}) -\mu(\hat{s}_{N(i)})-\\
  &  \displaystyle \sum\limits_{\hat{s}'_{N(i)}\in \bar{S}^{\mathrm{p}}_{N(i)}}\sum\limits_{\hat{\act}_{N(i)}\in\Act_{N(i)}}\probmdp_i^{\mathrm{p}}(\hat{s}_{N(i)}|\hat{s}'_{N(i)},\hat{\act}_{N(i)})o_i(s'_i,\hat{\act}_{N(i)}),\nonumber\\
\end{align*} 
i.e., the evaluation of the constraints~\eqref{eq:graph_mdp_flow} for agent $i$. 
Similarly, $g_3(o_i)$ and $g_4(o_i)$ of the evaluation of the constraints~\eqref{eq:graph_mdp_flow2} and~\eqref{eq:graph_mdp_spec}. 
$\mathcal{I}_0$ is the indicator function of $\lbrace 0 \rbrace$, i.e., $\mathcal{I}_0(\zeta)=0$ if $\zeta=0$, and $\mathcal{I}_0(\zeta)=\infty$ otherwise, and $\mathcal{I}_{+}$ is the indicator function of nonpositive reals, i.e., $\mathcal{I}_{+}(\zeta)=0$ if $\zeta\geq 0,$ and $\mathcal{I}_{+}(\zeta)=\infty$ otherwise~\cite[p.~218]{boyd_convex_optimization}.

\begin{algorithm}[t]
Initialize: $\displaystyle o^0_i$ and $\displaystyle\kappa^0_i~(i=1,2,\ldots, M)$\;
\For{$k=0,1,\ldots,I$}{
\For{$\displaystyle i=1,2,\ldots,M$}{
$\displaystyle z_i^{k+1} \gets - \frac{1}{M}\left(\sum\limits_{i=1}^M A_i o_i^k  - \frac{\kappa_i^k}{\beta} \right) + \left( A_i o_i^k  - \frac{\kappa_i^k}{\beta} \right).$
$\displaystyle o_i^{k+1} \gets\underset{o_i}{\text{argmin}}. f_i(o_i)+\frac{\beta}{2}\left\|A_io_i-z_i^{k+1}-\frac{\kappa^k_i}{\beta}\right\|_2^2.$

$\displaystyle\kappa_i^{k+1}\gets\kappa^k_i -\beta(A_i o_i^{k+1}-z^{k+1}_i).$
}
{
$\mathrm{res}_p\gets\sum_{i=1}^{M}\Vert A_io^k_i-z^k_i\Vert_2^2.$\\
$\mathrm{res}_d\gets\sum_{i=1}^{M}\beta\Vert \nu^k_i-\nu^{k-1}_i\Vert_2^2.$
}\\
\If{$\mathrm{res}_p\leq \gamma$ \textbf{ and } $\mathrm{res}_d\leq \gamma$}
   {\textbf{return} $o_i$ for $i=1,2, \ldots, M.$}}
\textbf{return} $o_i$ for $i=1,2, \ldots, M.$

\caption{Distributed Method with Primal Splitting ADMM} 
\label{Parallel_N}
\end{algorithm}

We introduce a set of auxiliary variables $z_i, i=1,\ldots,M$ and write the optimization problem in~\eqref{eq:multi_block_obj}--\eqref{eq:multi_block_cons} as following optimization problem

\begin{align}
\text{minimize} & \quad \displaystyle~\sum_{i=1}^M f_i(o_i)\label{eq:multi_block_obj1} \\
\text{subject to} &\quad \displaystyle~A_io_i =z_i,\quad i=1,\ldots,M,\label{eq:multi_block_cons1}\\
&\quad \displaystyle \sum_{i=1}^M z_i=0.\label{eq:multi_block_cons2}
\end{align}

The optimization problems in~\eqref{eq:multi_block_obj}--\eqref{eq:multi_block_cons} and in~\eqref{eq:multi_block_obj1}--\eqref{eq:multi_block_cons2} are equivalent in the sense that they share the same  optimal solution set for the variables $o_i, i=1,\ldots,M$, and the achieve the same optimal objective value.

We solve the optimization problem in a distributed manner by primal-splitting ADMM~\cite{wang2013solving}, which is given in Algorithm~\ref{Parallel_N}, where $\beta>0$ is the algorithm parameter, $\kappa_i$ is the dual variable for the constraints in~\eqref{eq:multi_block_cons1}, and $I$ is the number of iterations. The proposed method achieves an $O(1/k)$ convergence rate after $k$ iterations~\cite{wang2013solving}. 
The primal residual for the primal feasibility is given by $\mathrm{res}_p=\sum_{i=1}^{M}\Vert A_io^k_i-z^k_i\Vert_2^2$ and the dual residual for the dual feasibility is given by $\mathrm{res}_d =\sum_{i=1}^{M}\beta\Vert \nu^k_i-\nu^{k-1}_i\Vert_2^2$. 
The primal residual can be seen as the feasibility residual of the policy with respect to the specification, and the dual residual is the optimality residual of the policy with respect to the objective~\cite{boyd2011distributed}.
We stop the algorithm until it runs for $I$ iterations, or if the residuals are below a threshold $\gamma$.

\subsubsection{Complexity Analysis}
 
Computationally, the most expensive step of Algorithm 1 is to solve a LP for the $o_i$ update for $i=1,\ldots,M$. 
The number of variables and constraints in each LP for the $o_i$ update is exponential in $N(i)$, and each LP for the $o_i$ can be solved in time polynomial in the number of variables and constraints via interior-point methods~\cite{nesterov1994interior}.
Therefore, the computation time for each $o_i$ update is exponential in $N(i)$, and the computation time of the Algorithm 1 is linear in $M$. 

On the other hand, the number of variables and constraints in the optimization problem~\eqref{eq:multi_block_obj}--\eqref{eq:multi_block_cons} is linear in $M$ and exponential in $N(i)$. 
Therefore, if we solve the optimization problem in~\eqref{eq:multi_block_obj}--\eqref{eq:multi_block_cons} by an interior point method algorithm, then the overall complexity of the algorithm will be \emph{polynomial}, or typically cubic in $M$, and the computation will be challenging for a large number of agents. 
In our examples, we demonstrate that the running time for solving the optimization problem in~\eqref{eq:multi_block_obj}--\eqref{eq:multi_block_cons} directly does not scale linearly in $M$.

\section{Numerical Examples}

We demonstrate the proposed approach on two domains: (1) disease management in crop fields, (2) urban security. The simulations were performed on a computer with an Intel Core i5-7200u 2.50 GHz processor and 8 GB of RAM with Gurobi 9.0~\cite{gurobi} as the LP solver. 

\subsection{Disease Management in Crop Fields}
We consider the policy synthesis of a factored MDP for disease management in crop fields, which was introduced in~\cite{sabbadin2012framework}.
If a crop field is contaminated, it can infect its neighbors and the yield of that field decreases.
However, if a field is left fallow, it has a probability $\xi$ of recovering from contamination. 
The decisions of each year for each field $i$ include two actions ($\Act_i = \lbrace 1, 2\rbrace$) for each field: cultivate normally ($\act_i = 1$) or leave fallow ($\act_i = 2$). 

The problem is then to choose the optimal policy to maximize the expected yield. 
The topology of the fields is represented by an undirected graph, where each node in the graph represents one crop field.
An edge is drawn between two fields if the two fields share a common border and can infect each other. 
The number of neighbors for each field is four.
Each crop field can be in one of three states: $s_i = 1$ corresponds to the case where the field is uninfected. $s_i = 2$ and $s_i = 3$ correspond to different degrees of infection, with $s_i=3$ corresponding to a higher degree of infection. 
The probability that a field moves from state $s_i=1 $ to state $s_i =2$ or $s_i=2$ to state $s_i=3$ with $\act_i = 1$ is $\probmdp_i = \probmdp_i (\epsilon, p, n_i) = \epsilon + (1 - \epsilon) (1 - (1 - p)^{n_i} )$, where $\epsilon$ and $p$ are fixed parameters and $n_i$ is the number of the neighbors of $i$ that are infected. 
If the field $i$ is in $s_i=3$, then it remains in $s_i=3$ with a probability $1$ with $\act_i=1$.
The reward function depends on each field’s state and action. 
The maximal yield $r = 10$ is achieved by an uninfected, cultivated field, otherwise, the yield decreases linearly with the level of infection, from maximal reward $r$ to minimal reward $1 + r/10$. 
A field left fallow produces a reward of 1. 

\begin{table}[t]
\caption{The average yield with 100 fields with different values of $\lambda, p$ and $\xi$ with 50\% of the fields having a GTL specification.}
\label{table1}
\centering 

\begin{tabular}{|l|l|l|l|l|}
\hline
 \diagbox{$(p,\zeta)$}{$\lambda$}      & $0.9$ & $0.8$ & $0.7$ & $0.6$                  \\ \hline
$(0.1, 0.1)$        & $3.61$         & $ 4.15 $           & $ 4.57$              &       $5.05$                \\ \hline
$(0.2, 0.2)$        & $4.92$        & $ 5.81$          & $6.71$         & $7.13$           \\ \hline
$(0.5, 0.5)$        & $7.05$         & $8.12$       & $8.38$           & $8.43$           \\ \hline
$(0.8, 0.8)$        & $8.60$        & $8.95$          & $8.99$              & $9.00$              \\ \hline

\end{tabular}
\end{table}

Forster and Gilligan~\cite{forster2007optimizing} consider an optimal long-term disease control strategy of a crop field and determines that the optimal long-term control strategy requires treating most of the contaminated fields immediately to eradicate the disease.
Based on this control strategy, we consider the specification $\varphi=\neg \Diamond \Box_{\leq 2} d \wedge \neg \Diamond \Box_{\leq 3} \exists^{2} \bigcirc d$ for the critical fields, where $d$ means the field is infected.
A field satisfies the specification $\varphi$ if the field is never infected in two time steps in a row, and there should not be two neighboring fields that are infected three time steps in a row. 
This specification ensures that if a field or its neighboring field gets infected, it will be treated in a very short time to eradicate the disease and to prevent the disease from spreading, which is motivated by the control strategy of~\cite{forster2007optimizing}. 
Note that, in $\varphi$, the temporal parameter is $2$ for the critical fields, and is $3$ for the neighboring fields, as we intend to make sure that the critical fields are less infected than the non-critical fields, and also prevent the neighboring fields to spread the disease to the critical fields.

We consider the situation where 50\% of the fields are considered as critical fields, and we require that all the critical fields satisfy the specification $\varphi$ with a probability of at least $\lambda.$
Table~\ref{table1} shows the average yield for different values of $\lambda, p$ and $\zeta$, with $\epsilon=0.1$ and the parameter $\beta=1$ in Algorithm 1 after 500 iterations.
The average yield of the fields increases with a lower probability $\lambda$ of satisfying the specification $\varphi$. 
However, the fields may be contaminated with a higher probability with a lower probability $\lambda$ of satisfying the specification $\varphi$. 
For example, with a larger value of $\lambda$, e.g., $\lambda=0.9$, the critical fields (with a GTL specification) are infected on average with a probability of $0.03$, whereas the non-critical fields (without a GTL specification) are infected with a probability of $0.07$, showing that the critical fields are healthy with a higher probability. 
On the other hand, the probability of fields being healthy is lower with a lower threshold of $\lambda$, and the expected yield is higher.
This shows the trade-off between the average yield from the fields and the probability of the critical fields being uninfected.

\input{plots.tex}

In Figure~\ref{fig:residual}, we show the convergence rate of the approach with different values of the parameter $\beta$. 
The results in Figure~\ref{fig:residual} show that there is a clear trade-off between the magnitude of the parameter $\beta$ and the residuals during the iterations. 
With a higher value of $\beta$, e.g., $\beta=100$, the primal residual converges to a smaller value, e.g. to $10^{-4}$ after $500$ iterations. 
However, the dual residual only converges to a value of $10^{-2}$, and it is higher compared to the dual residuals with lower values of $\beta$.
We see that with the parameter selection of $\beta=1$, both the primal residual and the dual residual achieve relatively high accuracy, showing that the feasibility and optimality can be achieved with an accuracy of about $10^{-3}$ after $500$ iterations.
On the other hand, with a lower value of $\beta$, e.g., $\beta=0.01$, we obtain a lower dual residual, which means we obtain better performance in the objective value. 
However, the primal residual achieves lower accuracy with $\beta=0.01$.

We also show the scalability of the approach by varying the number of fields.  
We show the running time of the Algorithm~1 and the centralized method, which is solving the LP in~\eqref{eq:graph_mdp_obj}--\eqref{eq:graph_mdp_spec} directly, for different number of fields in Figure~\ref{fig:num_agent}. 
The results in Figure~\ref{fig:num_agent} shows that the running time of the approach scales linearly with the number of crop fields, and the centralized approach does not scale linearly with the number of crop fields. 
The Algorithm~1 is slower than the centralized method with fewer crop fields due to solving many smaller LP problems iteratively. 
However, with an increasing number of crop fields, the running time of the centralized method scales super-linearly and runs slower than Algorithm~1.

\begin{figure}[t]
\centering
\begin{tikzpicture}
\begin{axis}[%
width=2.417in,
height=1.0in, 
at={(2.167in,0.898in)},
scale only axis,
xmin=50,
xmax=1050,
yminorticks=true,
ymajorticks=true,
xmajorgrids=true,
ymajorgrids=true,
legend style={at={(0.0,1.05)},anchor=north west,font=\fontsize{6}{6.5}\selectfont},
xlabel style={font=\color{white!0!black}},
xlabel={Number of Crop Fields},
ymin=0,
ymax=400,
yminorticks=true,
ylabel style={font=\color{white!0!black}},
ylabel={Running Time},
axis background/.style={fill=white},
title style={font=\bfseries},
title={},
]
\addplot table[x=x,y=y] {data_agent.dat}; 
\addlegendentry{Distributed Method}
\addplot table[x=x,y=y] {data_agent2.dat}; 
\addlegendentry{Centralized Method}

\addplot [name path=upper1,draw=none] table[x=x,y expr=\thisrow{y}+\thisrow{ey}] {data_agent.dat};
\addplot [name path=lower1,draw=none] table[x=x,y expr=\thisrow{y}-\thisrow{ey}] {data_agent.dat};
\addplot [fill=blue!10] fill between[of=upper1 and lower1];
\addplot [name path=upper,draw=none] table[x=x,y expr=\thisrow{y}+\thisrow{ey}] {data_agent2.dat};
\addplot [name path=lower,draw=none] table[x=x,y expr=\thisrow{y}-\thisrow{ey}] {data_agent2.dat};
\addplot [fill=red!10] fill between[of=upper and lower];
\end{axis}
\end{tikzpicture}
\caption{The running time of distributed method and centralized method with a different number of crop fields after 500 iterations with the distributed method. The shaded region shows the standard deviation of the running times in 20 runs. The decentralized algorithm scales linearly with the number of crop fields. The centralized method is slower than the Algorithm~1 with an increasing number of crop fields and does not scale linearly.}
\label{fig:num_agent}
\end{figure}

\subsection{Urban Security}

We consider an urban security problem, where a criminal plans his next move randomly based on the information on the nearby locations that are protected by police patrols~\cite{wu2019reward}. 
There are $M$ police officers assigned to monitor the locations. 
Each police officer coordinates with a sub-group of police officers in monitoring to prevent crimes.

Figure~\ref{fig:agent_distribution} shows $35$ intersections in San Francisco, CA with $7$ rows and $5$ columns~\cite{wu2019reward}.
We use a factored MDP to describe the network of the set of intersections.
We show the number of crimes that occurred in October 2018 within 500 feet of each intersection in Figure~\ref{fig:agent_distribution}\footnote{The crime data can be found here https://www.crimemapping.com/map/ca/sanfrancisco.}.
A police officer obtains a reward that is equal to the number of crimes in an intersection if the police officer monitors that intersection.
In this way, we incentivize the police officers to monitor intersections with higher crime rates.

\begin{figure}[t] 
    \centering
    \input{test.tex}
    \caption{The $35$ intersections in the northeast of San Francisco. The map is obtained from Google maps. The numbering of the states starts from the bottom left corner and goes from left to right in every row. The number beside each location indicates the number of crimes in that area. We denote the critical intersections by labeling time with a \textquotedblleft$*$\textquotedblright~next to the number of crimes. The average number of police officers monitoring each intersection is given by different colors.}
    \label{fig:agent_distribution}
\end{figure}

On the other hand, there exists a set of critical intersections (see the intersections with \textquotedblleft$*$\textquotedblright~next to the number of crimes in Figure~\ref{fig:agent_distribution}).
To ensure that the critical intersections are monitored often by a police officer, we consider a GTL specification for the critical intersections.
Specifically, for each critical intersection $s_{\textrm{crit}}$, we assign a police officer $i$ with the specification $\varphi = \Box \Diamond_{\leq 3} (s_{\textrm{crit}} \vee  \exists^{1} \bigcirc s_{\textrm{crit}})$. 
The specification $\varphi$ means that each critical intersection $s_{\textrm{crit}}$ should be visited at least once in every three time steps by either the police officer $i$ or a neighboring police officer of $i$.
This specification means that the critical intersections are monitored sufficiently often by a police officer, and it also aims to prevent the police officers from monitoring the intersections with the highest crime rates at all times.

We consider an example with 15 police officers, where each police officer is responsible for 9 intersections on a $3 \times 3$ grid. 
For example, a police officer is responsible for intersections between $5\leq x_2 \leq 7$ and $3 \leq x_1 \leq 5$, and another police officer is responsible for intersections between $5\leq x_2 \leq 7$ and $2 \leq x_1 \leq 4$. 
The overall objective of the police officers is to maximize the expected reward by monitoring states with the highest crime rates while satisfying the GTL specification $\varphi$. 

For $\lambda=0.9$, Figure~\ref{fig:agent_distribution} shows the obtained assignment for the police officers with $\beta=1$ after 500 iterations. 
The computation time of the approach was 164.3 seconds.  
Different colors at each intersection show the average number of police officers monitoring that intersection for a given hour. 
The results show the GTL specification $\varphi$ is satisfied by the obtained assignment.
We also observe that the police officers monitor the intersections with higher crime rates (critical or non-critical) to maximize the expected reward while satisfying the GTL specification $\varphi$.
Without the GTL specification, the police officers would monitor the intersections with the highest crime rate to obtain maximal expected reward, while ignoring the critical intersections. 
By enforcing the GTL specification, the results in Figure~\ref{fig:agent_distribution} shows that the critical intersections are visited sufficiently often (satisfying $\varphi$), while the intersections with higher crime rates are visited as often as possible to maximize the expected reward.   

\section{Conclusions and Future Work} 

We proposed a method for the distributed synthesis of policies for multi-agent systems to implement spatial-temporal tasks. 
We express the spatial-temporal tasks in a specification language called graph temporal logic.
For such systems, we decomposed the synthesis problem into a set of smaller synthesis problems, one for each agent. 
With our numerical examples, we showed that the algorithm runs in time linear in the number of agents and scales to hundreds of agents.

For future work, we will extend the framework to allow the edges of the graph to be time-varying.
We will consider scenarios where the edges of the graph are functions of the agent's local states, and the agents may share the state information with its neighboring agents in certain parts of the state spaces. 
For example, the agents may share their state information when they are close to each other and need to coordinate with each other to achieve a task.

\bibliographystyle{ACM-Reference-Format}  
\bibliography{sample-bibliography}  

\end{document}

%% file: plots.tex
\begin{figure}[t]
\centering
\definecolor{mycolor1}{rgb}{0.00000,0.44700,0.74100}%
\definecolor{mycolor2}{rgb}{0.85000,0.32500,0.09800}%
\definecolor{mycolor3}{rgb}{0.92900,0.69400,0.12500}%
\definecolor{mycolor4}{rgb}{0.49400,0.18400,0.55600}%
\definecolor{mycolor5}{rgb}{0.46600,0.67400,0.18800}%
\definecolor{mycolor6}{rgb}{0.30100,0.74500,0.93300}%
\begin{tikzpicture}

\begin{axis}[
width=2.517in,
height=2.3in,
at={(2.167in,0.898in)},
scale only axis,
log basis y={10},
legend style={at={(0.5,0.60)},anchor=south west,font=\fontsize{6}{6.5}\selectfont},
tick align=outside,
tick pos=both,
x grid style={lightgray!92.026!black},
xmin=-24.95, xmax=523.95,
yminorticks=true,
ymajorticks=true,
xmajorgrids=true,
ymajorgrids=true,
xtick style={color=black},
y grid style={lightgray!92.02!black},
ylabel={some numbers},
ymin=0.000003898, ymax=50.232,
ymode=log,
ytick style={color=black},
xlabel style={font=\color{white!0!black}},
xlabel={Number of iterations},
ylabel style={font=\color{white!0!black}},
ylabel={Norm of Residuals},
axis background/.style={fill=white},
legend style={legend cell align=left, align=left, draw=white!15!black},
] 
\addplot [color=mycolor5,line width=1.5pt] table[x=x,y=y] {primal_res_001.dat};
\addlegendentry{Primal Residual, $\beta=0.01$}
\addplot [color=mycolor6,line width=1.5pt] table[x=x,y=y] {dual_res_001.dat};
\addlegendentry{Dual Residual, $\beta=0.01$}
\addplot [color=mycolor1,dashed, line width=1.5pt] table[x=x,y=y] {primal_res_1.dat};
\addlegendentry{Primal Residual, $\beta=1$}
\addplot [color=mycolor2, dashed,line width=1.5pt] table[x=x,y=y] {dual_res_1.dat};
\addlegendentry{Dual Residual, $\beta=1$}
\addplot [color=mycolor3,dashdotted,line width=1.5pt] table[x=x,y=y] {primal_res_100.dat};
\addlegendentry{Primal Residual, $\beta=100$}
\addplot [color=mycolor4,dashdotted,line width=1.5pt] table[x=x,y=y] {dual_res_100.dat};
\addlegendentry{Dual Residual, $\beta=100$}

\end{axis} 

\end{tikzpicture}
\caption{Norm of primal residual and dual residual versus the number of iterations for the crop fields problem with different values of the parameter $\beta$.}
\label{fig:residual}
\end{figure}

%% file: test.tex
\begin{tikzpicture}
\centering
\begin{axis}[
xmin=0, xmax=5.5,
ymin=0.5, ymax=7.5,
xlabel={$x_1$},ylabel={$y_2$},
width=9.0cm,height=8.7cm,colorbar horizontal,point meta min=0.0,
    point meta max=1.1,colorbar style={
        width=6.5cm,
        xtick={0,0.2,0.4,0.6,0.8,1,1.2}}
]
\addplot graphics [includegraphics cmd=\pgfimage,xmin=0, xmax=5.5, ymin=0.5, ymax=7.5] {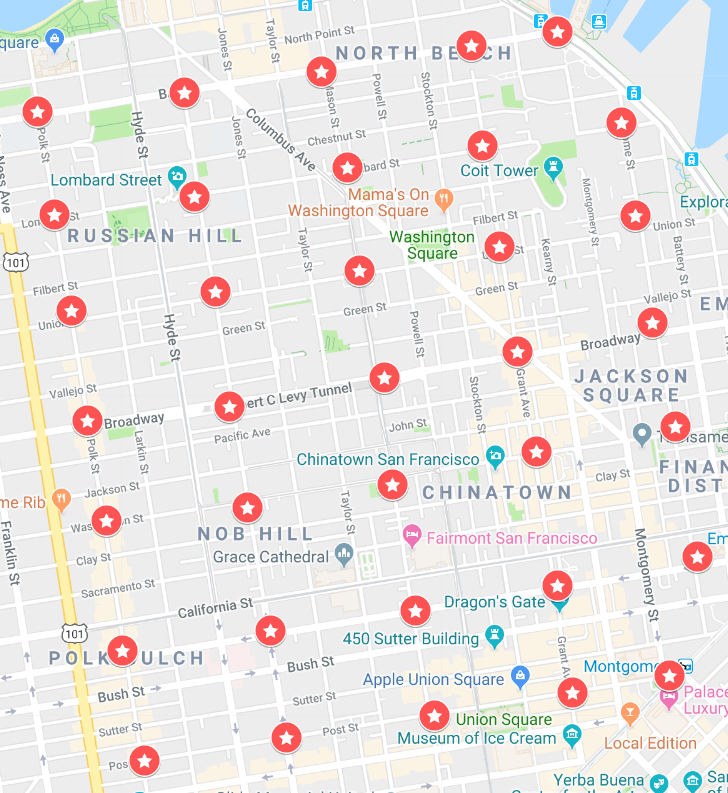};
\addplot  [only marks, mark size=0.26cm, fill opacity=0.9, draw opacity=0.9, scatter, scatter src=explicit]
table [x=x, y=y, meta=colordata]{%
x                      y                      colordata
0.2865 6.51 0.312142070174261
1.394 6.673 0.685085927040432
2.431 6.855 0.914646248532035
3.563 7.085 0.510038617619258
4.217 7.206 0.268309670841333
0.420 5.595 0.0312786093990236
1.475 5.754 0.3733020503831
2.63 6.02 0.647700301868446
3.655 6.205 0.0956116641490765
4.703 6.406 0.292572508784498
0.55 4.76 0.192803655791449
1.645 4.92 0.190913094921226
2.725 5.095 0.224785003737408
3.783 5.315 0.111976513627961
4.805 5.575 0.219931620010259
0.664 3.788 0.431622108593821
1.738 3.90 0.396370445602672
2.91 4.155 0.147413625198577
3.915 4.387 0.506321106674979
4.941 4.6605 0.22862013464441
0.805 2.895 0.389295244783753
1.875 3.01 0.248893015403111
2.973 3.22 0.527589436383504
4.06 3.50 0.942273497490569
5.105 3.73 0.432355433055291
0.93 1.75 0.512572727714095
2.045 1.935 0.573277350638577
3.14 2.11 0.302103086335453
4.215 2.325 0.584277457601353
5.27 2.58 0.334918815894527
1.095 0.78 0.403516938521447
2.165 0.99 0.548737505681162
3.29 1.19 1.1184029105688
4.33 1.38 0.660896781604525
5.06 1.53 0.639444820729708
};
\node [draw=none] at (110,70) {$\textbf{17}$};
\node [draw=none] at (215,90) {$\textbf{25}$};
\node [draw=none] at (327,119) {$\textbf{48}^{\displaystyle \bf{*}}$};
\node [draw=none] at (433,140) {$\textbf{69}$};
\node [draw=none] at (505,155) {$\textbf{80}$};
\node [draw=none] at (92,180) {$\textbf{43}$};
\node [draw=none] at (203,194) {$\textbf{19}$};
\node [draw=none] at (313,214) {$\textbf{15}$};
\node [draw=none] at (420,234) {$\textbf{23}$};
\node [draw=none] at (526,258) {$\textbf{21}$};
\node [draw=none] at (79,288) {$\textbf{17}$};
\node [draw=none] at (185,300) {$\textbf{9}$};
\node [draw=none] at (296,321) {$\textbf{2}$};
\node [draw=none] at (407,351) {$\textbf{36}^{\displaystyle \bf{*}}$};
\node [draw=none] at (512,372) {$\textbf{8}$};
\node [draw=none] at (65,378) {$\textbf{17}$};
\node [draw=none] at (174,393) {$\textbf{19}$};
\node [draw=none] at (288,412) {$\textbf{2}$};
\node [draw=none] at (392,439) {$\textbf{11}$};
\node [draw=none] at (493,468) {$\textbf{8}$};
\node [draw=none] at (54,469) {$\textbf{5}$};
\node [draw=none] at (163,490) {$\textbf{7}$};
\node [draw=none] at (272,507) {$\textbf{10}$};
\node [draw=none] at (378,527) {$\textbf{12}$};
\node [draw=none] at (482,551) {$\textbf{7}$};
\node [draw=none] at (41,555) {$\textbf{2}$};
\node [draw=none] at (146,568) {$\textbf{25}$};
\node [draw=none] at (262,598) {$\textbf{7}$};
\node [draw=none] at (366,617) {$\textbf{5}$};
\node [draw=none] at (470,642) {$\textbf{12}$};
\node [draw=none] at (29,652) {$\textbf{17}$};
\node [draw=none] at (139,662) {$\textbf{26}$};
\node [draw=none] at (244,683) {$\textbf{33}^{\displaystyle \bf{*}}$};
\node [draw=none] at (326,679) {$\textbf{23}$};
\node [draw=none] at (457,684) {$\textbf{14}$};
\end{axis}

\end{tikzpicture}